\PassOptionsToPackage{unicode}{hyperref}
\PassOptionsToPackage{hyphens}{url}
\PassOptionsToPackage{dvipsnames,svgnames,x11names}{xcolor}
\documentclass[
  12pt]{article}
\usepackage{xcolor}
\usepackage{amsmath,amssymb}
\usepackage{iftex}
\ifPDFTeX
  \usepackage[T1]{fontenc}
  \usepackage[utf8]{inputenc}
  \usepackage{textcomp} 
\else 
  \usepackage{unicode-math} 
  \defaultfontfeatures{Scale=MatchLowercase}
  \defaultfontfeatures[\rmfamily]{Ligatures=TeX,Scale=1}
\fi
\usepackage{lmodern}
\ifPDFTeX\else
\fi
\IfFileExists{upquote.sty}{\usepackage{upquote}}{}
\IfFileExists{microtype.sty}{
  \usepackage[]{microtype}
  \UseMicrotypeSet[protrusion]{basicmath} 
}{}
\makeatletter
\@ifundefined{KOMAClassName}{
  \IfFileExists{parskip.sty}{%
    \usepackage{parskip}
  }{
    \setlength{\parindent}{0pt}
    \setlength{\parskip}{6pt plus 2pt minus 1pt}}
}{
  \KOMAoptions{parskip=half}}
\makeatother
\makeatletter
\ifx\paragraph\undefined\else
  \let\oldparagraph\paragraph
  \renewcommand{\paragraph}{
    \@ifstar
      \xxxParagraphStar
      \xxxParagraphNoStar
  }
  \newcommand{\xxxParagraphStar}[1]{\oldparagraph*{#1}\mbox{}}
  \newcommand{\xxxParagraphNoStar}[1]{\oldparagraph{#1}\mbox{}}
\fi
\ifx\subparagraph\undefined\else
  \let\oldsubparagraph\subparagraph
  \renewcommand{\subparagraph}{
    \@ifstar
      \xxxSubParagraphStar
      \xxxSubParagraphNoStar
  }
  \newcommand{\xxxSubParagraphStar}[1]{\oldsubparagraph*{#1}\mbox{}}
  \newcommand{\xxxSubParagraphNoStar}[1]{\oldsubparagraph{#1}\mbox{}}
\fi
\makeatother

\usepackage{longtable,booktabs,array}
\usepackage{calc} 
\usepackage{etoolbox}
\makeatletter
\patchcmd\longtable{\par}{\if@noskipsec\mbox{}\fi\par}{}{}
\makeatother
\IfFileExists{footnotehyper.sty}{\usepackage{footnotehyper}}{\usepackage{footnote}}
\makesavenoteenv{longtable}
\usepackage{graphicx}
\makeatletter
\newsavebox\pandoc@box
\newcommand*\pandocbounded[1]{
  \sbox\pandoc@box{#1}%
  \Gscale@div\@tempa{\textheight}{\dimexpr\ht\pandoc@box+\dp\pandoc@box\relax}%
  \Gscale@div\@tempb{\linewidth}{\wd\pandoc@box}%
  \ifdim\@tempb\p@<\@tempa\p@\let\@tempa\@tempb\fi
  \ifdim\@tempa\p@<\p@\scalebox{\@tempa}{\usebox\pandoc@box}%
  \else\usebox{\pandoc@box}%
  \fi%
}
\def\fps@figure{htbp}
\makeatother

\providecommand{\tightlist}{%
  \setlength{\itemsep}{0pt}\setlength{\parskip}{0pt}}

\usepackage[]{natbib}
\usepackage{booktabs}
\usepackage{longtable}
\usepackage{array}
\usepackage{multirow}
\usepackage{wrapfig}
\usepackage{float}
\usepackage{colortbl}
\usepackage{pdflscape}
\usepackage{tabu}
\usepackage{threeparttable}
\usepackage{threeparttablex}
\usepackage[normalem]{ulem}
\usepackage{makecell}
\usepackage{xcolor}
\usepackage{amsmath}
\usepackage{float}
\usepackage{hyperref}
\usepackage[utf8]{inputenc}
\usepackage{bm}
\def\tightlist{}
\usepackage{setspace}
\newcommand\pD{$p\text{-}D$}

\newcommand\gD{$2\text{-}D$}

\makeatletter
\@ifpackageloaded{caption}{}{\usepackage{caption}}
\AtBeginDocument{%
\ifdefined\contentsname
  \renewcommand*\contentsname{Table of contents}
\else
  \newcommand\contentsname{Table of contents}
\fi
\ifdefined\listfigurename
  \renewcommand*\listfigurename{List of Figures}
\else
  \newcommand\listfigurename{List of Figures}
\fi
\ifdefined\listtablename
  \renewcommand*\listtablename{List of Tables}
\else
  \newcommand\listtablename{List of Tables}
\fi
\ifdefined\figurename
  \renewcommand*\figurename{Figure}
\else
  \newcommand\figurename{Figure}
\fi
\ifdefined\tablename
  \renewcommand*\tablename{Table}
\else
  \newcommand\tablename{Table}
\fi
}
\@ifpackageloaded{float}{}{\usepackage{float}}
\floatstyle{ruled}
\@ifundefined{c@chapter}{\newfloat{codelisting}{h}{lop}}{\newfloat{codelisting}{h}{lop}[chapter]}
\floatname{codelisting}{Listing}

\makeatother
\makeatletter
\@ifpackageloaded{caption}{}{\usepackage{caption}}
\@ifpackageloaded{subcaption}{}{\usepackage{subcaption}}
\makeatother
\usepackage{bookmark}
\IfFileExists{xurl.sty}{\usepackage{xurl}}{} 
\hypersetup{
  pdftitle={Choosing Better NLDR Layouts by Evaluating the Model in the High-dimensional Data Space},
  pdfauthor={Jayani P. Gamage; Dianne Cook; Paul Harrison; Michael Lydeamore; Thiyanga S. Talagala},
  pdfkeywords={high-dimensional data, dimension reduction, data
visualization, statistical graphics, tour, unsupervised learning},
  colorlinks=true,
  linkcolor={blue},
  filecolor={Maroon},
  citecolor={Blue},
  urlcolor={Blue},
  pdfcreator={LaTeX via pandoc}}

\begin{document}

\def\spacingset#1{\renewcommand{\baselinestretch}%
{#1}\small\normalsize} \spacingset{1}


\title{\bf Choosing Better NLDR Layouts by Evaluating the Model in the
High-dimensional Data Space}
\author{
Jayani P. Gamage\\
Econometrics \& Business Statistics, Monash University\\
and\\Dianne Cook\\
Econometrics \& Business Statistics, Monash University\\
and\\Paul Harrison\\
MGBP, BDInstitute, Monash University\\
and\\Michael Lydeamore\\
Econometrics \& Business Statistics, Monash University\\
and\\Thiyanga S. Talagala\\
Statistics, University of Sri Jayewardenepura\\
}
\maketitle

\bigskip
\bigskip
\begin{abstract}
Nonlinear dimension reduction (NLDR) techniques such as tSNE, and UMAP
provide a low-dimensional representation of high-dimensional data
(\pD{}) by applying a nonlinear transformation. NLDR often exaggerates
random patterns. But NLDR views have an important role in data analysis
because, if done well, they provide a concise visual (and conceptual)
summary of \pD{} distributions. The NLDR methods and hyper-parameter
choices can create wildly different representations, making it difficult
to decide which is best, or whether any or all are accurate or
misleading. To help assess the NLDR and decide on which, if any, is the
most reasonable representation of the structure(s) present in the \pD{}
data, we have developed an algorithm to show the \gD{} NLDR model in the
\pD{} space, viewed with a tour, a movie of linear projections. From
this, one can see if the model fits everywhere, or better in some
subspaces, or completely mismatches the data. Also, we can see how
different methods may have similar summaries or quirks.
\end{abstract}

\noindent%
{\it Keywords:} high-dimensional data, dimension reduction, data
visualization, statistical graphics, tour, unsupervised learning
\vfill

\newpage
\spacingset{1.9} 

\section{Introduction}\label{introduction}

Nonlinear dimension reduction (NLDR) is popular for making a convenient
low-dimensional (\(k\text{-}D\)) representation of high-dimensional
(\(p\text{-}D\)) data (\(k < p\)). Recently developed methods include
t-distributed stochastic neighbor embedding (tSNE) \citep{laurens2008},
uniform manifold approximation and projection (UMAP) \citep{leland2018},
potential of heat-diffusion for affinity-based trajectory embedding
(PHATE) algorithm \citep{moon2019}, large-scale dimensionality reduction
using triplets (TriMAP) \citep{amid2022}, and pairwise controlled
manifold approximation (PaCMAP) \citep{yingfan2021}.

However, the representation generated can vary dramatically from method
to method, choice of hyper-parameter or even random seed, as illustrated
by Figure~\ref{fig-NLDR-variety}. The specific method and
hyper-parameters used to produce each layout (see Supplementary
Materials) is not essential for the discussion. The dilemma for the
analyst is which representation to use. The choice might result in
different procedures used in the downstream analysis, or different
inferential conclusions. Various academics have expressed concerns with
current practices and procedures for choosing (e.g.
\citet{irizarry2024}, \citet{chari2023}). The research described here
provides new numerical and visual tools to aid with this decision.

\begin{figure}

\centering{

\includegraphics[width=0.8\linewidth,height=\textheight,keepaspectratio]{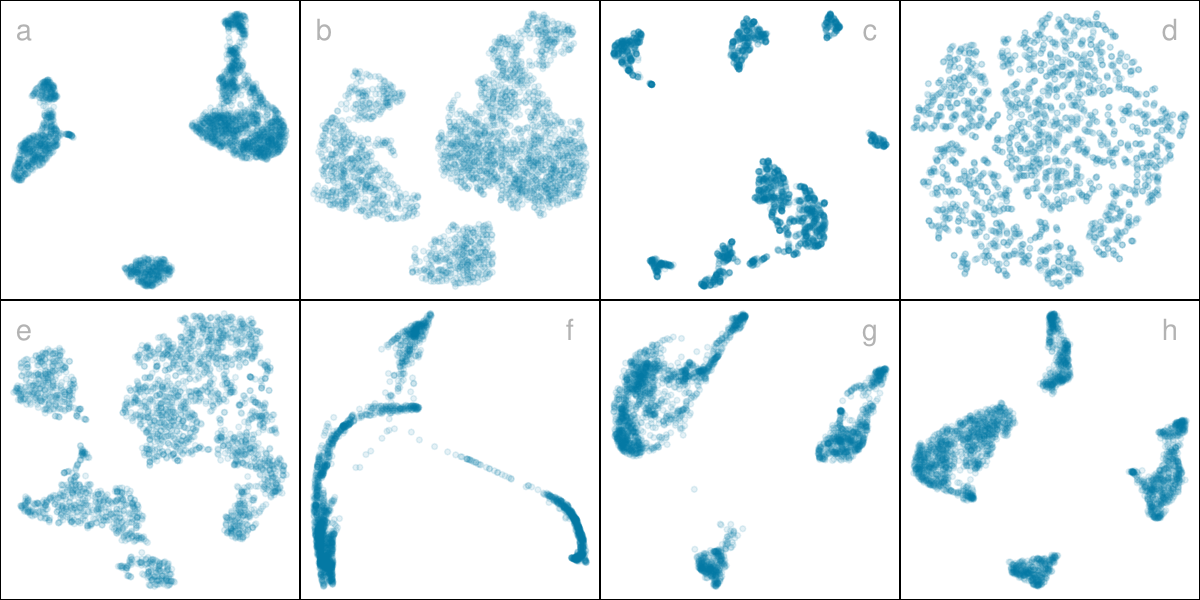}

}

\caption{\label{fig-NLDR-variety}Eight different NLDR representations of
the same data, produced by tSNE, UMAP, PHATE, TriMAP and PaCMAP with a
variety of hyper-parameter choices. The variety in layouts makes it
difficult to choose which best represents the data distribution.}

\end{figure}%

The paper is organized as follows. Section~\ref{sec-background-visalgo}
provides a summary of the literature on NLDR, and high-dimensional data
visualization methods. Section~\ref{sec-method-visalgo} contains the
details of the new methodology, including a simulated data example. In
Section~\ref{sec-bestfit-visalgo}, we describe how to assess the best
fit and identify the most accurate \(2\text{-}D\) layout based on the
proposed model diagnostics. Two applications illustrating the use of the
new methodology for bioinformatics and image classification are in
Section~\ref{sec-applications-visalgo}. Limitations and future
directions are provided in Section~\ref{sec-discussion-visalgo}.

\section{Background}\label{sec-background-visalgo}

Historically, low-dimensional (\(k\text{-}D\)) representations of
high-dimensional (\(p\text{-}D\)) data have been computed using
multidimensional scaling (MDS) \citep{kruskal1964}, which includes
principal components analysis (PCA) (for an overview see
\citet{jolliffe2011}). (A contemporary comprehensive guide to MDS can be
found in \citet{borg2005}.) The \(k\text{-}D\) representation can be
considered to be a layout of points in \(k\text{-}D\) produced by an
embedding procedure that maps the data from \(p\text{-}D\). In MDS, the
\(k\text{-}D\) layout is constructed by minimizing a stress function
that differences distances between points in \(p\text{-}D\) with
potential distances between points in \(k\text{-}D\). Various
formulations of the stress function result in non-metric scaling
\citep{saeed2018} and isomap \citep{silva2002}. Challenges in working
with high-dimensional data, including visualization, are outlined in
\citet{johnstone2009}.

Many new methods for NLDR have emerged in recent years, all designed to
better capture specific structures potentially existing in
\(p\text{-}D\). Here we focus on five currently popular techniques:
tSNE, UMAP, PHATE, TriMAP and PaCMAP. The methods tSNE, UMAP, TriMAP and
PaCMAP can be considered for producing the \(k\text{-}D\) representation
by minimizing the divergence between two inter-point distance
distributions. PHATE is an example of a diffusion process spreading to
capture geometric shapes, that include both global and local structure.
(See \citet{coifman2005} for an explanation of diffusion processes.)

The array of layouts in Figure~\ref{fig-NLDR-variety} illustrate what
can emerge from the choices of method and hyper-parameters, and the
random seed that initiates the computation. Key structures interpreted
from these views suggest: (1) highly \textbf{separated clusters} (a, b,
e, g, h) with the number ranging from 3-6; (2) \textbf{stringy branches}
(f), and (3) \textbf{barely separated clusters} (c, d) which would
\textbf{contradict} the other representations. These contradictions
arise because these methods and hyper-parameter choices provide
different lenses on the interpoint distances in the data.

The alternative approach to visualizing the high-dimensional data is to
use linear projections. PCA is the classical approach, resulting in a
set of new variables which are linear combinations of the original
variables. Tours, defined by \citet{As85}, broaden the scope by
providing movies of linear projections, that provide views the data from
all directions. (See \citet{lee2021} for a review of tour methods.)
There are many tour algorithms implemented, with many available in the R
package \texttt{tourr} \citep{wickham2011}, and versions enabling better
interactivity in \texttt{langevitour} \citep{harisson2024} and
\texttt{detourr} \citep{hart2022}. Linear projections are a safe way to
view high-dimensional data, because they do not warp the space, so they
are more faithful representations of the structure. However, linear
projections can be cluttered, and global patterns can obscure local
structure. The simple activity of projecting data from \(p\text{-}D\)
suffers from piling \citep{laa2022}, where data concentrates in the
center of projections. NLDR is designed to escape these issues, to
exaggerate structure so that it can be observed. But as a result NLDR
can hallucinate wildly, to suggest patterns that are not actually
present in the data.

Our proposed solution is to use the tour to examine how the NLDR is
warping the space. It follows what \citet{wickham2015} describes as
\emph{model-in-the-data-space}. The fitted model should be overlaid on
the data, to examine the fit relative the spread of the observations.
While this is straightforward, and commonly done when data is
\(2\text{-}D\), it is also possible in \(p\text{-}D\), for many models,
when a tour is used.

\citet{wickham2015} provides several examples of models overlaid on the
data in \(p\text{-}D\). In hierarchical clustering, a representation of
the dendrogrom using points and lines can be constructed by augmenting
the data with points marking merging of clusters. Showing the movie of
linear projections reveals shows how the algorithm sequentially fitted
the cluster model to the data. For linear discriminant analysis or
model-based clustering the model can be indicated by \((p-1)\text{-}D\)
ellipses. It is possible to see whether the elliptical shapes
appropriately matches the variance of the relevant clusters, and to
compare and contrast different fits. For PCA, one can display the model
(a \(k\text{-}D\) plane of the reduced dimension) using wireframes of
transformed cubes. Using a wireframe is the approach we take here, to
represent the NLDR model in \(p\text{-}D\).

\section{Method}\label{sec-method-visalgo}

\subsection{What is the NLDR model?}\label{what-is-the-nldr-model}

At first glance, thinking of NLDR as a modeling technique might seem
strange. It is a simplified representation or abstraction of a system,
process, or phenomenon in the real world. The \(p\text{-}D\)
observations are the realization of the phenomenon, and the
\(k\text{-}D\) NLDR layout is the simplified representation. Typically,
\(k=2\), which is used for the rest of this paper. From a statistical
perspective we can consider the distances between points in the
\(2\text{-}D\) layout to be variance that the model explains, and the
(relative) difference with their distances in \(p\text{-}D\) is the
error, or unexplained variance. We can also imagine that the positioning
of points in \(2\text{-}D\) represent the fitted values, that will have
some prescribed position in \(p\text{-}D\) that can be compared with
their observed values. This is the conceptual framework underlying the
more formal versions of factor analysis \citep{joreskog1969} and MDS.
(Note that, for this thinking the full \(p\text{-}D\) data needs to be
available, not just the interpoint distances.)

We define the NLDR as a function
\(g\text{:}~ \mathbb{R}^{n\times p} \rightarrow \mathbb{R}^{n\times 2}\),
with hyper-parameters \(\bm{\theta}\). These parameters,
\(\bm{\theta}\), depend on the choice of \(g\), and can be
considered part of model fitting in the traditional sense. Common
choices for \(g\) include functions used in tSNE, UMAP, PHATE, TriMAP,
PaCMAP, or MDS, although in theory any function that does this mapping
is suitable.

\begin{figure}

\centering{

\includegraphics[width=1\linewidth,height=\textheight,keepaspectratio]{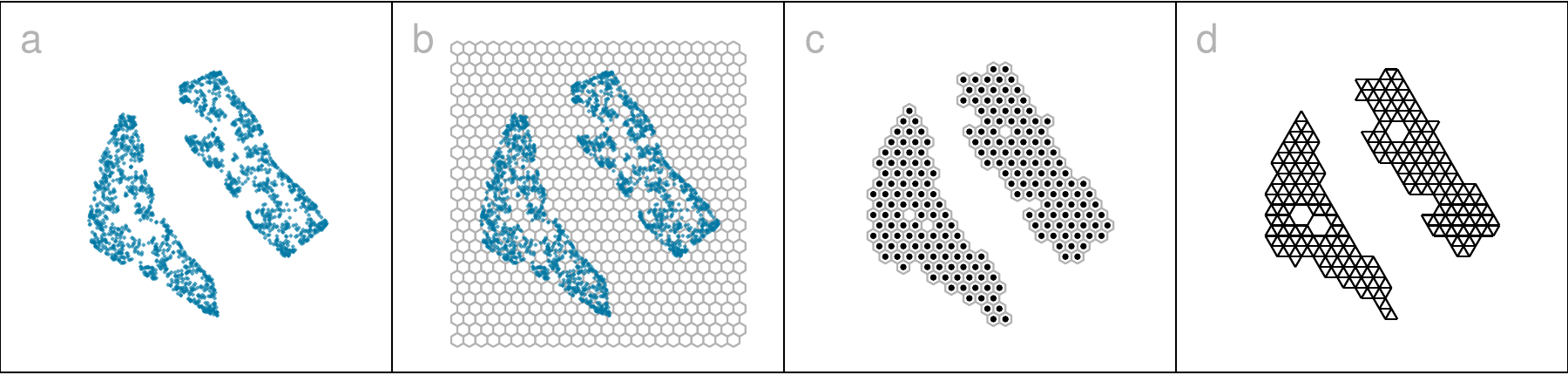}

}

\caption{\label{fig-NLDR-two-curvy}Key steps for constructing the model
on the tSNE layout (\(k=2\)) of 2NC7: (a) data, (b) hexagon bins, (c)
bin centroids, and (d) triangulated centroids. The 2NC7 data is shown.}

\end{figure}

With our goal being to make a representation of this \(2\text{-}D\)
layout that can be lifted into high-dimensional space, the layout needs
to be augmented to include neighbor information. A simple approach would
be to triangulate the points and add edges. A more stable approach is to
first bin the data, reducing it from \(n\) to \(m\leq n\) observations,
and connect the bin centroids. We recommend using a hexagon grid because
it better reflects the data distribution and has less artifacts than a
rectangular grid. This process serves to reduce some noisiness in the
resulting surface shown in \(p\text{-}D\). The steps in this process are
shown in Figure~\ref{fig-NLDR-two-curvy}, and documented below.

To illustrate the method and how to use it to choose a reasonable
layout, we use \(7\text{-}D\) simulated data, which we call the ``2NC7''
data. It has two separated nonlinear clusters, one forming a
\(2\text{-}D\) curved shape, and the other a \(3\text{-}D\) curved
shape, each consisting of \(1000\) observations. The first four
variables hold this cluster structure, and the remaining three are
purely noise. We would consider \(T=(X_1, X_2, X_3, X_4)\) to be the
geometric structure (true model) that we hope to capture. This data is
sufficiently simple, with just two complexities (two separated
curvilinear clusters, and two different implicit dimensions), to
adequately explain the new method. The applications section contains two
practical examples where NLDR has been used in published work. This data
has both global and local structure. The two separated clusters would be
considered to be global structure, and the nonlinear low-dimensional
shapes could be considered to be local structure, one being
\(2\text{-}D\) and the other \(3\text{-}D\). An ideal NLDR layout would
reveal the two clusters with moderate separation, and flatten the
curvilinear forms while preserving the proximity of points.

\subsection{\texorpdfstring{Algorithm to represent the model in
\(2\text{-}D\)}{Algorithm to represent the model in 2\textbackslash text\{-\}D}}\label{algorithm-to-represent-the-model-in-2text-d}

\subsubsection{Scale the data}\label{scale-the-data}

Because we are working with distances between points, starting with data
having a standard scale, e.g.~{[}0, 1{]}, is recommended. The default
should take the aspect ratio produced by the NLDR
\((r_1, r_2, ..., r_k)\) into account. When \(k=2\), as in hexagon
binning, the default range is \([0, y_{i,\text{max}}], i=1,2\), where
\(y_{1,\text{max}}=1\) and \(y_{2,\text{max}} = r_2/r_1\)
(Figure~\ref{fig-NLDR-two-curvy}). If the NLDR aspect ratio is ignored
then set \(y_ {2,\text{max}} = 1.\)

\subsubsection{Hexagon grid
configuration}\label{hexagon-grid-configuration}

\begin{figure}

\centering{

\includegraphics[width=0.3\linewidth,height=\textheight,keepaspectratio]{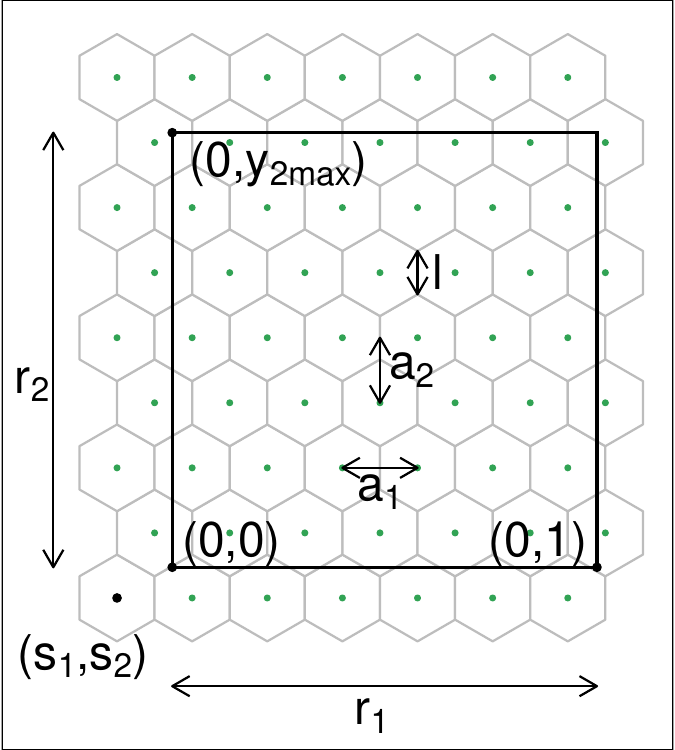}

}

\caption{\label{fig-hex-param}The components of the hexagon grid
illustrating notation.}

\end{figure}

Although there are several implementations of hexagon binning
\citep{carr1987}, and a published paper \citep{dan2023}, surprisingly,
none has sufficient detail or components that produce everything needed
for this project. So we described the process used here.
Figure~\ref{fig-hex-param} illustrates the notation used.

The \(2\text{-}D\) hexagon grid is defined by its bin centroids. Each
hexagon, \(H_h\) (\(h = 1, \dots, b\)) is uniquely described by
centroid, \(C_{h}^{(2)} = (c_{h1}, c_{h2})\). The number of bins in each
direction is denoted as \((b_1, b_2)\), with \(b = b_1 \times b_2\)
being the total number of bins. We expect the user to provide just
\(b_1\) and we calculate \(b_2\) using the NLDR ratio, to compute the
grid.

To ensure that the grid covers the range of data values a buffer
parameter (\(q\)) is set as a proportion of the range. By default,
\(q=0.1\). The buffer should be extending a full hexagon width (\(a_1\))
and height (\(a_2\)) beyond the data, in all directions. The lower left
position where the grid starts is defined as \((s_1, s_2)\), and
corresponds to the centroid of the lowest left hexagon,
\(C_{1}^{(2)} = (c_{11}, c_{12})\). This must be smaller than the
minimum data value. Because it is one buffer unit, \(q\) below the
minimum data values, \(s_1 = -q\) and \(s_2 = -qr_2\).

The value for \(b_2\) is computed by fixing \(b_1\). Considering the
upper bound of the first NLDR component, \(a_1 > (1+2q)/(b_1 -1)\).
Similarly, for the second NLDR component,

\[
a_2 \geq \frac{r_2 + q(1 + r_2)}{(b_2 - 1)}.
\]

Since \(a_2 = \sqrt{3}a_1/2\) for regular hexagons,

\[
a_1 \geq \frac{2[r_2 + q(1 + r_2)]}{\sqrt{3}(b_2 - 1)}.
\]

This is a linear optimization problem. Therefore, the optimal solution
must occur on a vertex. Therefore,

\begin{equation}\phantomsection\label{eq-bin2}{
b_2 = \Big\lceil1 +\frac{2[r_2 + q(1 + r_2)](b_1 - 1)}{\sqrt{3}(1 + 2q)}\Big\rceil.
}\end{equation}

\subsubsection{Binning the data}\label{binning-the-data}

Observations are grouped into bins based on their nearest centroid. This
produces a reduction in size of the data from \(n\) to \(m\), where
\(m\leq b\) (total number of bins). This can be defined using the
function
\(u: \mathbb{R}^{n\times 2} \rightarrow \mathbb{R}^{m\times 2}\), where
\[u(i) = \arg\min_{j = 1, \dots, b} \sqrt{(y_{i1} - C^{(2)}_{j1})^2 + (y_{i2} - C^{(2)}_{j2})^2},\]
maps observation \(i\) into \(H_h = \{i| u(i) = h\}\).

By default, the bin centroid is used for describing a hexagon (as done
in Figure~\ref{fig-NLDR-two-curvy} (c)), but any measure of center, such
as a mean or weighted mean of the points within each hexagon, could be
used. The bin centers, and the binned data, are the two important
components needed to render the model representation in high dimensions.

\subsubsection{Indicating neighborhood}\label{indicating-neighborhood}

Delaunay triangulation \citep{lee1980, alb2024} is used to connect
points so that edges indicate neighboring observations, in both the NLDR
layout (Figure~\ref{fig-NLDR-two-curvy} (d)) and the \(p\text{-}D\)
model representation. When the data has been binned the triangulation
connects centroids. The edges preserve the neighborhood information from
the \(2\text{-}D\) representation when the model is lifted into
\(p\text{-}D\).

\subsection{\texorpdfstring{Rendering the model in
\(p\text{-}D\)}{Rendering the model in p\textbackslash text\{-\}D}}\label{rendering-the-model-in-ptext-d}

The last step is to lift the \(2\text{-}D\) model into \(p\text{-}D\) by
computing \(p\text{-}D\) vectors that represent bin centroids. We use
the \(p\text{-}D\) mean of the points in a given hexagon, \(H_h\),
denoted \(C_{h}^{(p)}\), to map the centroid
\(C_{h}^{(2)} = (c_{h1}, c_{h2})\) to a point in \(p\text{-}D\). Let the
\(j^{th}\) component of the \(p\text{-}D\) mean be

\[C_{hj}^{(p)} = \frac{1}{n_h}\sum_{i =1}^{n_h} x_{hij}, ~~~h = 1, \dots, b;~ j=1, \dots, p; ~ n_h > 0.\]

\begin{figure}[!ht]

\centering{

\includegraphics[width=1\linewidth,height=\textheight,keepaspectratio]{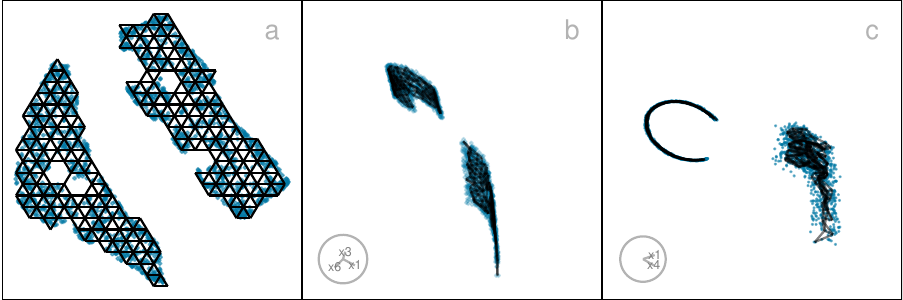}

}

\caption{\label{fig-best-fit-tsne}Lifting the \(2\text{-}D\) fitted
model into \(p\text{-}D\). Two projections of the \(p\text{-}D\) fitted
model overlaying the data are shown in b, c.~The fit is reasonably tight
with the data in one cluster (top one in b), but slightly less so in the
other cluster probably because it is \(3\text{-}D\). Notice also that,
in the \(2\text{-}D\) layout the two clusters have internal gaps which
creates a model with some holes. This lacy pattern happens regardless of
the hyper-parameter choice, but this doesn't severely impact the
\(p\text{-}D\) model representation.}

\end{figure}%

\subsection{Measuring the fit}\label{sec-summary}

All NLDR methods internally optimize a quantity to produce a layout for
any particular hyper-parameter set. These are not always made available
in the model output, and may not be universally comparable between
hyper-parameter choices and methods.

Several common metrics are often used to assess the quality of any NLDR
layout, based on preservation of global and local structure of the data.
The \(RNX\) curve quantifies the neighborhood agreement between
\(p\text{-}D\) and \(k\text{-}D\) spaces, by computing the area under
the curve (\(ARNX\)) across a range of neighborhood scales
\citep{john2015}. A high value indicates better preservation of a
balance of global and local structure. Random Triplet Accuracy (RTA) and
Centroid Triplet Accuracy compare the order of \(2\text{-}D\) and
\(p\text{-}D\) distances of random triplets of points
\citep{yingfan2021}. High values indicate preservation of the geometry,
suggesting both local and global structure preservation. The Shepard
diagram \citep{shepard1962} and its associated Spearman correlation (SC)
\citep{spearman1961} between \(p\text{-}D\) and \(k\text{-}D\)
distances. High values indicate preservation of global structure. The
Global Score (GS) measures how well an embedding retains the overall
geometry of the data relative to a PCA baseline \citep{amid2022}. Higher
values indicate better preservation of global structure. The metric RTA,
SC, GS, and ARNX have been reversed (rRTA, rSC, rGS, and rARNX) so that
they align with HBE - the lower the value the better the layout.

\begin{figure}

\centering{

\includegraphics[width=1\linewidth,height=\textheight,keepaspectratio]{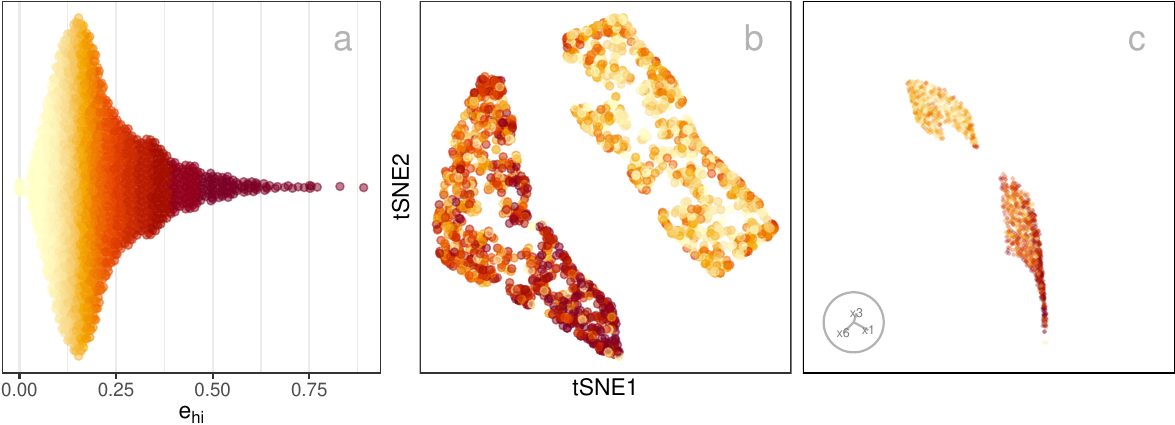}

}

\caption{\label{fig-p-d-error-in-2d-two-curvy}Examining the distribution
of residuals in a jittered dotplot (a), \(2\text{-}D\) NLDR layout (b)
and a tour of \(4\text{-}D\) data space (c). Color indicates residual
(\(e_{hi}\)), dark color indicating high value. Most large residuals are
distributed in one cluster (bottom one in c) and most small residuals
are distributed in the other cluster.}

\end{figure}

None of the above measures is particularly well-suited to assessing our
model fit, as we will show later. Thus, we need a different approach to
measuring model fit. Because the model here is similar to a confirmatory
factor analysis model (see a general explanation in \citet{brown2015}),
\(\widehat{T} + \epsilon\), our approach is similar to the ones used in
this area. it is based on ``residuals'' computed as the difference
between the fitted model and observed values in \(p\text{-}D\).
Observations are associated with their bin center, \(C_{h}^{(p)}\),
which are also considered to be the \emph{fitted values}. In factor
analysis language, these fitted values might also be denoted as
\(\widehat{X}\). The error is computed by taking the squared
\(p\text{-}D\) Euclidean distance, of points from their bin centroid,
which we will call hexbin error (HBE):

\begin{equation}\phantomsection\label{eq-equation1}{ HBE = \sqrt{\frac{1}{n}\sum_{h = 1}^{m}\sum_{i = 1}^{n_h}\sum_{j = 1}^{p} (\bm{x}_{hij} - C^{(p)}_{hj})^2}}\end{equation}

where \(n\) is the number of observations, \(m\) is the number of
non-empty bins, \(n_h\) is the number of observations in \(h^{th}\) bin,
\(p\) is the number of variables and \(\bm{x}_{hij}\) is the
\(j^{th}\) dimensional data of \(i^{th}\) observation in \(h^{th}\)
hexagon. We can consider
\(e_{hi} = \sqrt{\sum_{j = 1}^{p} (\bm{x}_{hij} - C^{(p)}_{hj})^2}\)
to be the residual for each observation.
Figure~\ref{fig-p-d-error-in-2d-two-curvy} shows plots of \(e\) as a
density (a), coloring the points in the NLDR layout (b) and the points
in a tour (c). It can be see that the biggest residuals are in one
cluster, which occurs due the intentional design that one cluster is
slightly \(3\text{-}D\) and perfectly captured by a \(2\text{-}D\)
layout.

\subsection{\texorpdfstring{Prediction into
\(2\text{-}D\)}{Prediction into 2\textbackslash text\{-\}D}}\label{prediction-into-2text-d}

NLDR methods are primarily designed for visualization and exploration
rather than reconstruction, and do not explicitly provide out-of-sample
prediction. Of the five methods studied here, only UMAP provides a
\texttt{predict()} function for embedding new data points based on the
learned manifold \citep{tomasz2023}. Several other approaches, not used
here, PCA, neural network autoencoders \citep{hinton2006} and parametric
tSNE \citep{van2009} support prediction.

A benefit of our approach is that for any NLDR method, it provides a way
to predict the layout position of a new observation, \(x'\). The steps
are (1) determine the closest bin centroid in \(p\text{-}D\),
\(C^{(p)}_{h}\) and (2) predict the embedding to be the bin centroid in
\(2\text{-}D\), \(C^{(2)}_{h}\).

\subsection{Tuning}\label{tuning}

The model fitting is based on several parameters, including the hexagon
bin parameters and the low count bin removal process. The hexagon bin
parameters define the bottom-left bin position \((s_1, \ s_2)\), the
number of bins in the horizontal direction (\(b_1\)), which also
determines the number of bins in the vertical direction (\(b_2\)), the
total number of bins (\(b\)), and the total number of non-empty bins
(\(m\)). Low count bins are removed using standardized bin counts,
defined as \(w_h = n_h/n, ~~h=1, \dots m\).

Default values are provided for each of these, but deciding on the best
model fit is assisted by examining a range of values. The default number
of bins \(b=b_1\times b_2\) is computed based on the sample size, by
setting \(b_1=n^{1/3}\), consistent with the Diaconis-Freedman rule
\citep{freedman1981}. The value of \(b_2\) is determined analytically by
\(b_1, q, r_2\) (Equation~\ref{eq-bin2}). Values of \(b_1\) between
\(2\) and \(b_1 = \sqrt{n/r_2}\) are recommended, where the dependence
on \(r_2\) reflects the preservation of aspect ratio in the NLDR layout.

\begin{figure}

\centering{

\includegraphics[width=1\linewidth,height=\textheight,keepaspectratio]{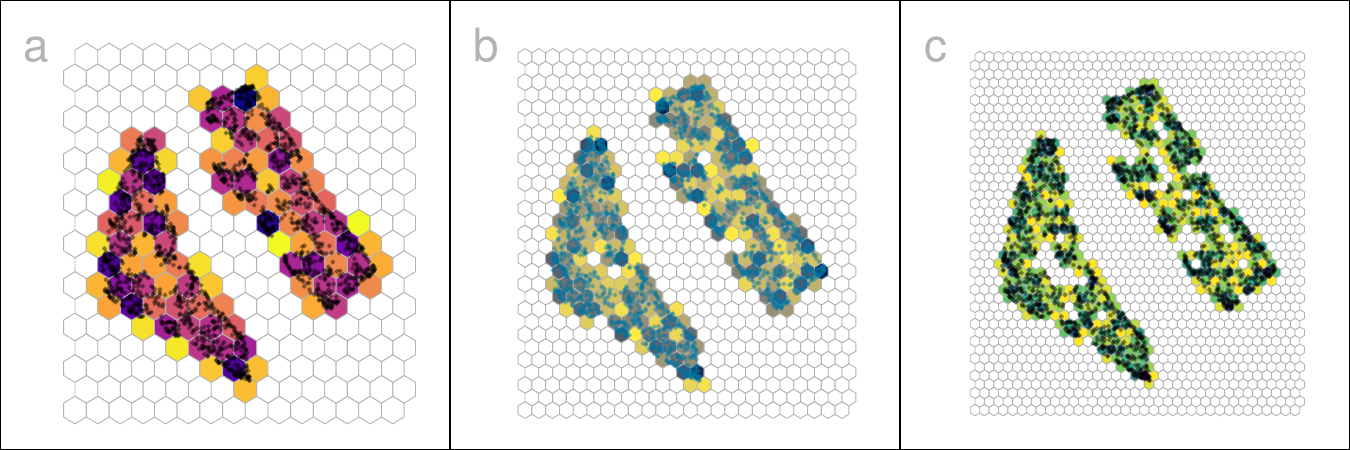}

}

\caption{\label{fig-bins-two-curvy}Hexbin density plots of tSNE layout
of the 2NC7 data, using three different \(b_1\) specifications yielding
different \(b_2, b, m\): (a) \textbf{15}, 18, 270, 98, (b) \textbf{24},
29, 696, 209, and (c) \textbf{35}, 42, 1470, 386. Color indicates
standardized counts, dark indicating high count and light indicates low
count. At the smallest binwidth, the data structure is discontinuous,
suggesting that there are too many bins.}

\end{figure}

Figure~\ref{fig-bins-two-curvy} shows the hexbin grids for three choices
of \(b_1\). While the number of bins is the common parameter to modify,
bin start positions \((s_1, \ s_2)\) can be worth experimenting with
also because it can also change bin counts.

\begin{figure}

\centering{

\includegraphics[width=1\linewidth,height=\textheight,keepaspectratio]{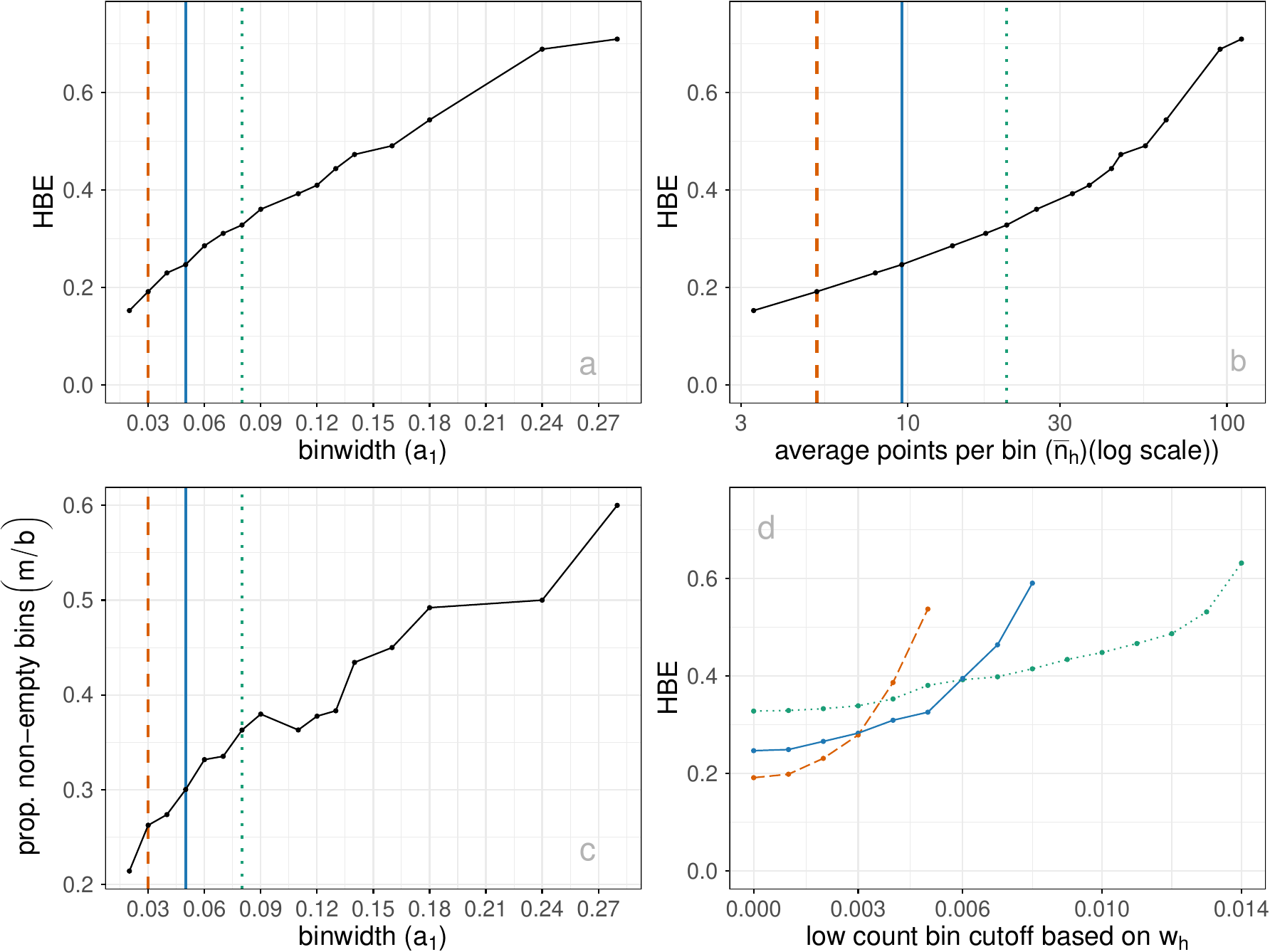}

}

\caption{\label{fig-param-two-curvy}Various plots to help tune the model
fit: (a) HBE vs \(a_1\), (b) HBE vs \(\bar{n}_h\), (c) proportion of
non-empty bins (\(m/b\)) vs \(a_1\), (d) HBE vs \(w_h\) cutoff used for
removing low count bins. Color indicates the three binwidths show in
Figure 6: \(0.03\) (orange dashed), \(0.05\) (blue solid), and \(0.08\)
(green dotted). The better model fit will have low HBE, but reasonably
sized bins that capture the data sufficiently. Proportion of non-empty
bins tends to increase with \(a_1\) (c). Removing a few low count bins
doesn't substantially change HBE, for all three binwidths (d).}

\end{figure}

It is worthwhile to consider what are desirable aspects of a hexbin
result, that maps to summarizing the \(p\text{-}D\) fit well. The
binning should capture the underlying data distribution closely, with
minimum number of necessary bins. An ideal binning might be indicated by
a more uniform distribution of bin counts, or having few relatively
empty bins. To help with this assessment average bin count
(\(\bar{n} = \sum{n_h}/m\)), average standardized bin count
(\(\bar{w} = \sum{w_h}/m\)) and proportion of non-empty bins (\(m/b\)),
are also computed. Figure~\ref{fig-param-two-curvy} shows some choices
of plots of these quantities for a single NLDR layout, with three
choices of \(a_1\) indicated. Some expectations and reasoning for these
plots is:

\begin{itemize}
\tightlist
\item
  HBE will increase as \(a_1\) increases, so good choices will be just
  before a big increase. In plot a, HBE changes fairly steadily so there
  is no easy choice to make.
\item
  HBE can also be examined against average standardized bin count or
  average bin count (plot b). This is similar to the comparison with
  \(a_1\) but to use when comparing different NLDR layouts. Different
  layouts might produce different density of points, which will not be
  captured well by a comparison of HBE vs \(a_1\).
\item
  The proportion of non-empty bins is interesting to examine across
  different binwidths (plot c). A good binning should have just the
  right amount of bins to neatly cover the shape of the data, and no
  more or less. As binwidth gets smaller, \(m/b\) should roughly get
  bigger.
\item
  Bins with a small number of observations might be removed to sharpen
  the wireframe model. This can have adverse effects, though - failing
  to extend the wireframe into sparse areas, or resulting in holes in
  the wireframe. Plot d shows the relationship between HBE (computed for
  all observations despite some bin removal) and the standardized bin
  count cutoff used to remove bins. For all three chosen bin widths a
  small number of bins can be removed without affecting HBE.
\end{itemize}

\subsection{Interactive graphics}\label{interactive-graphics}

Matching points in the \(2\text{-}D\) layout with their positions in
\(p\text{-}D\) is useful when tuning the fit. This can be used to
examine the fitted model in some subspaces in \(p\text{-}D\), in
particular in association with residual plots.

The interactive \(2\text{-}D\) layout \citep{chapman2020} and the
langevitour \citep{harisson2024} view with the fitted model overlaid can
be linked using a browsable HTML widget (\citet{joe2025},
\citet{joe2024}). A rectangular ``brush'' is used to select points in
one plot, which will highlight the corresponding points in the other
plot(s). Because the langevitour is dynamic, brush events that become
active will pause the animation, so that a user can interrogate the
current view. This approach will be illustrated on the examples, to show
how it can help to understand how the NLDR has organized the
observations, and learn where it does not do well.

\section{\texorpdfstring{Choosing the best \(2\text{-}D\)
layout}{Choosing the best 2\textbackslash text\{-\}D layout}}\label{sec-bestfit-visalgo}

\begin{figure}

\centering{

\includegraphics[width=1\linewidth,height=\textheight,keepaspectratio]{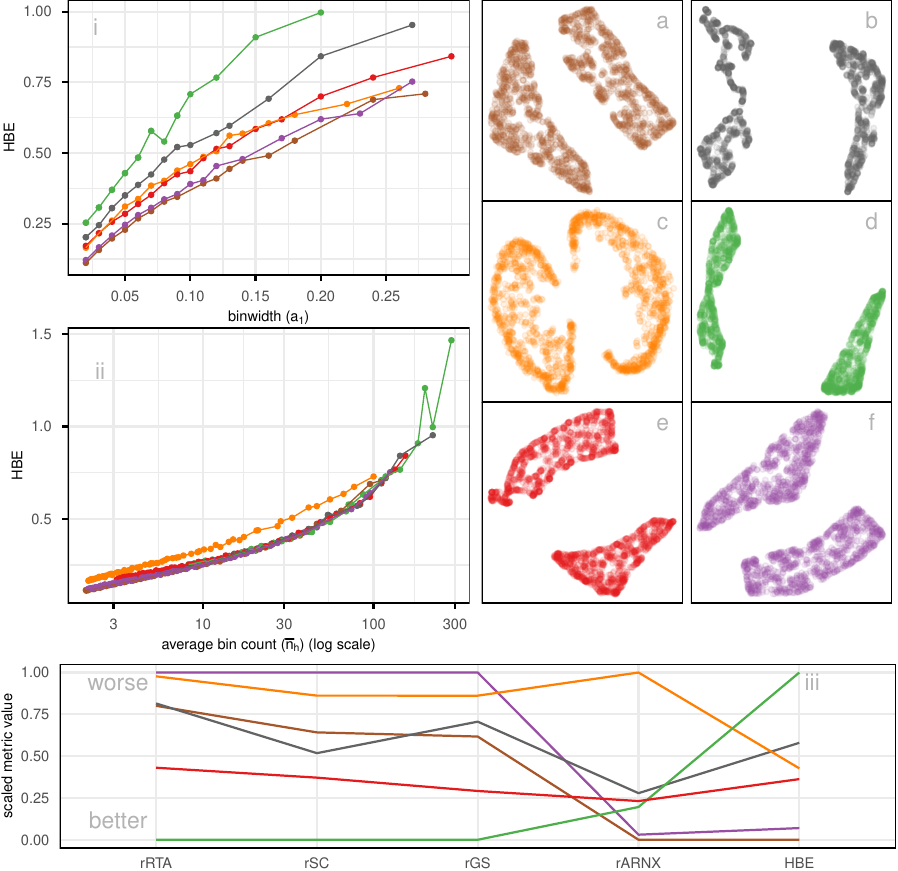}

}

\caption{\label{fig-toy-rmse}Assessing which of the 6 NLDR layouts (a-f)
on the 2NC7 data is the better representation using HBE for varying (i)
binwidth (\(a_1\)), and (ii) average bin count (\(\bar{n}_h\)). Color
represents NLDR layout. Layout d is universally poor. Layouts b, e that
show two close clusters are universally suboptimal. Layout f with little
separation performs well at tiny binwidth (where most points are in
their own bin) and poorly as binwidth increases. Layout e has small
separation with oddly shaped clusters. Layout a is the best choice. Plot
(ii), which compares HBE values with respect to average bin count, helps
account for differences in cluster density; here, the variation among
layouts is reduced, showing that some differences observed in (i) arise
from density rather than true structure. Comparison of scaled evaluation
metrics (rRTA, rSC, rGS, rARNX, and HBE using \(a_1=0.05\)) for the six
NLDR layouts computed on the 2NC7 data using a parallel coordinate plot
(iii). Color of the line indicates NLDR layout.}

\end{figure}

Figure~\ref{fig-toy-rmse} illustrates the approach to compare the fits
for different representations and assess the strength of any fit. What
does it mean to be a best fit for this problem? Analysts use an NLDR
layout to display the structure present in high-dimensional data in a
convenient \(2\text{-}D\) display. It is a competitor to linear
dimension reduction that can better represent nonlinear associations
such as clusters. However, these methods can hallucinate, suggesting
patterns that don't exist, and grossly exaggerate other patterns. Having
a layout that best fits the high-dimensional structure is desirable but
more important is to identify bad representations so they can be
avoided. The goal is to help users decide on a the most useful and
appropriate low-dimensional representation of the high-dimensional data.

A particular pattern that we commonly see is that analysts tend to pick
layouts with clusters that have big separations between them. When you
examine their data in a tour, it is almost always that we see there are
no big separations, and actually often the suggested clusters are not
even present. While we don't expect that analysts include animated gifs
of tours in their papers, we should expect that any \(2\text{-}D\)
representation adequately indicates the clustering that is present, and
honestly show lack of separation or lack of clustering when it doesn't
exist. It is important for analysts to have tools to select the accurate
representation not the pretty but wrong representation.

To decide on a reasonable layout, an analyst needs a selection of NLDR
representations generated using a range of hyper-parameter choices and
possibly different methods, such as tSNE and UMAP. They also require a
range of model fits created by varying the binwidths and the level of
low count bin removal, along with the calculated HBE values for each
layout after transformation into the \(p\text{-}D\) space. Finally, the
analyst must be able to visually examine how well each model fits the
data in the original data space.

Comparing the HBE to obtain the best fit is appropriate if the same NLDR
method is used. However, because the HBE is computed on \(p\text{-}D\)
data it measures the fit between model and data so it can also be used
to compare the fit of different NLDR methods. A lower HBE indicates a
better NLDR representation.

Figure~\ref{fig-toy-rmse} compares the metrics rARNX, rRTA, rSC, rGS,
along with HBE computed on \(a_1=0.05\) for the six layouts shown in
Figure~\ref{fig-toy-rmse}. This is a parallel coordinate plot where the
y-axis shows a normalized score to ensure the metrics are on the same
scale. Each line corresponds to one layout.

There is some agreement between the metrics. All, except rARNX and HBE
agree that layout d is best. rARNX and HBE agree that layout f is best
or very close to best. Layout a is best according to HBE and rARNX but
considered to be much less optimal by rRTA, rSC and rGS. Layout c is
considered poor by rRTA, rSC, rGS, and rARNX. This illustrates how
difficult it is to use the numerical metrics alone to decide on the best
layout.

The problem with rSC is that correlation is not a good measure in the
presence of clusters - the further the clusters are apart in the layout
produces a Shepard plot with two clusters of distances which will
produce a high correlation value. Similar reasoning would explain why
rRTA and rGS behave similarly: they put too much emphasis on the global
structure. Thus, for the 2NC7 data further apart clusters score better,
overly emphasizing that there are two clusters, even though this
separation is not accurately reflecting the difference in
\(p\text{-}D\).

\begin{figure}

\centering{

\includegraphics[width=1\linewidth,height=\textheight,keepaspectratio]{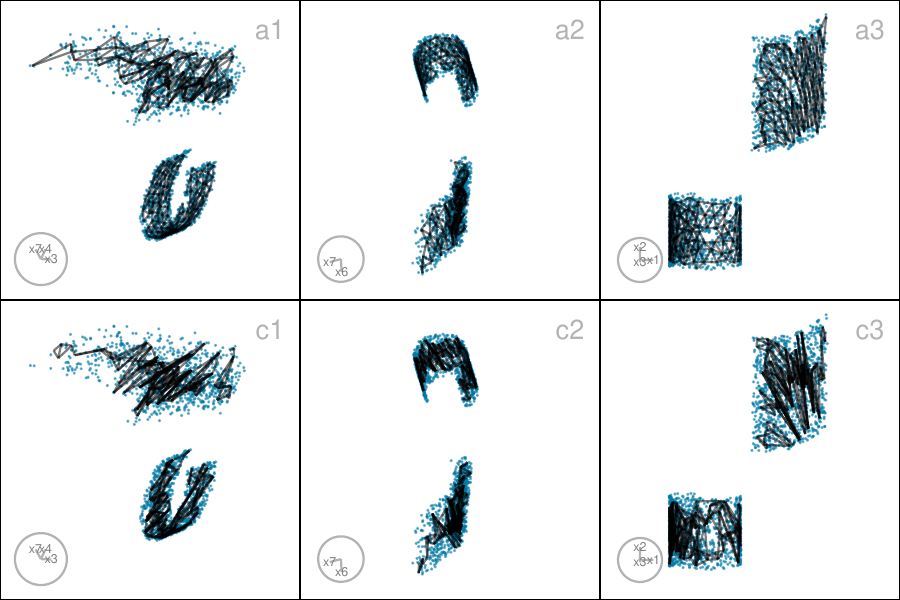}

}

\caption{\label{fig-fit-tsne-phate}Three \(2\text{-}D\) projections from
a tour showing the fitted models (black lines) for layouts a (top row)
and c (bottom row) of the 2NC7 data (blue points). Layout c, which was
poorly rated by rARNX and HBE covers less of the width of the data than
a. The triangular gridding is less visible in c than layout a, and
actually corresponds to extreme twisting.}

\end{figure}

When the metrics disagree is causes confusion for the analyst, and thus
provides a temptation to choose the nicest looking layout (very
separated clusters), even though it may be a hallucination. Because HBE
is accompanied with a representation of the layout in \(p\text{-}D\) to
compare with the observed data, it can help to add more clarity in
making decisions. Figure~\ref{fig-fit-tsne-phate} shows the fitted
models for layouts a (rated high by HBE and rARNX) and c (rated poorly).
These are \(2\text{-}D\) projections from the tour, with black
indicating the fitted model overlaid on the blue points of the data. The
reason for the poor fit is that the PHATE layout (c) twists extremely
along the 2- and \(3\text{-}D\) manifolds where the data lies. We have
learned that all the NLDR methods tend to have twists in the fit in
\(p\text{-}D\) but this is extreme. This is likely why layout c has poor
metrics relative to the other layouts, and it suggests that it does not
adequately capture the local structure in the 2NC7 data.

\section{Applications}\label{sec-applications-visalgo}

To illustrate the approach we use two examples: PBMC3k data (single cell
gene expression) where an NLDR layout is used to represent cluster
structure present in the \(p\text{-}D\) data, and MNIST hand-written
digits where NLDR is used to represent essentially a low-dimensional
nonlinear manifold in \(p\text{-}D\).

\subsection{PBMC3k}\label{sec-pbmc}

This is a benchmark single-cell RNA-Seq data set collected on Human
Peripheral Blood Mononuclear Cells (PBMC3k) as used in \citet{pbmc2019}.
Single-cell data measures the gene expression of individual cells in a
sample of tissue (see for example, \citet{haque2017}). This type of data
is used to obtain an understanding of cellular level behavior and
heterogeneity in their activity. Clustering of single-cell data is used
to identify groups of cells with similar expression profiles. NLDR is
often used to summarize the cluster structure. Usually, NLDR does not
use the cluster labels to compute the layout, but uses color to
represent the cluster labels when it is plotted.

In this data there are \(2622\) single cells and \(1000\) gene
expressions (variables). Following the same pre-processing as
\citet{chen2024}, different NLDR techniques were performed on the first
nine principal components. Figure~\ref{fig-NLDR-variety} shows this data
using a variety of methods, and different hyper-parameters. You can see
that the result is wildly different depending on the choices. Layout a
is a reproduction of the layout that was published in \citet{chen2024}.
This layout suggests that the data has three very well separated
clusters, each with an odd shape. The question is whether this
accurately represents the cluster structure in the data, or whether they
should have chosen b or c or d or e or f or g or h. This is what our new
method can help with -- to decide which is the more accurate
\(2\text{-}D\) representation of the cluster structure in the
\(p\text{-}D\) data.

\begin{figure}

\centering{

\includegraphics[width=1\linewidth,height=\textheight,keepaspectratio]{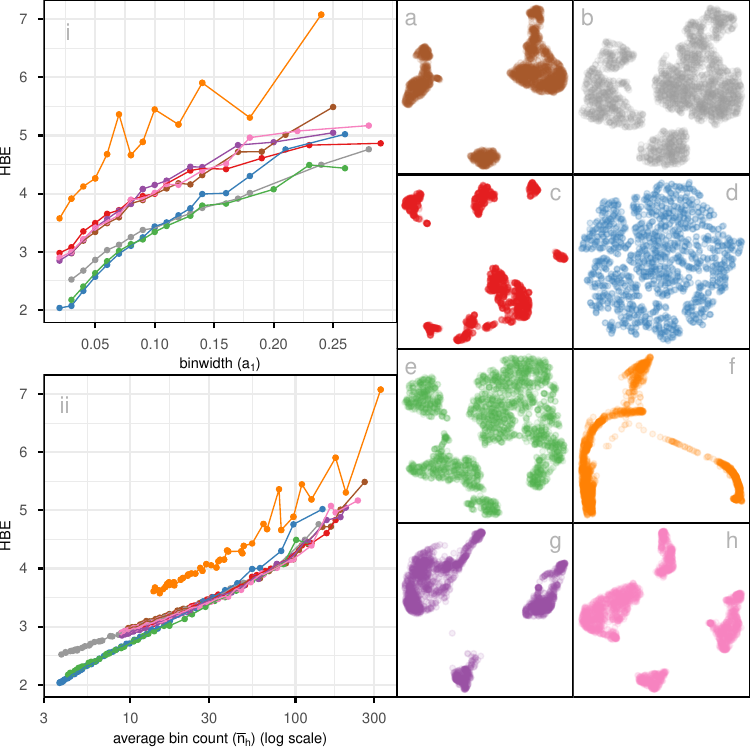}

}

\caption{\label{fig-pbmc-rmse}Assessing which of the 8 NLDR layouts on
the PBMC3k data (shown in Figure~\ref{fig-NLDR-variety}) is the better
representation using HBE for varying (i) binwidth (\(a_1\)), and (ii)
average bin count (\(\bar{n}_h\)). Color used for the lines and points
in the left plot and in the scatterplots represents NLDR layout (a-h).
Layout f is universally poor. Layouts a, c, g, h that show large
separations between clusters are universally suboptimal. Layout d with
little separation performs well at tiny binwidth (where most points are
in their own bin) and poorly as binwidth increases. The choice of best
is between layouts b and e, that have small separations between oddly
shaped clusters. Layout e is chosen as the best. Plot (ii), which
accounts for the density within clusters by using average bin count,
shows reduced differences between layouts, indicating that part of the
variation in (i) is driven by cluster density rather than true
structural differences.}

\end{figure}

Figure~\ref{fig-pbmc-rmse} shows HBE across a range of binwidths
(\(a_1\)) for each of the layouts in Figure~\ref{fig-NLDR-variety}. The
layouts were generated using tSNE and UMAP with various hyper-parameter
settings, while PHATE, PaCMAP, and TriMAP were applied using their
default settings. Lines are color coded to match the color of the
layouts shown on the right. Lower HBE indicates the better fit. Using a
range of binwidths shows how the model changes, with possibly the best
model being one that is universally low HBE across all binwidths. It can
be seen that layout f is sub-optimal with universally higher HBE. Layout
a, the published one, is better but it is not as good as layouts b, d,
or e. With some imagination layout d perhaps shows three barely
distinguishable clusters. Layout e shows three, possibly four, clusters
that are more separated. The choice reduces from eight to these two.
Layout d has slightly better HBE when the \(a_1\) is small, but layout e
beats it at larger values. Thus we could argue that layout e is the most
accurate representation of the cluster structure, of these eight.

\begin{figure}[!ht]

\centering{

\includegraphics[width=0.9\linewidth,height=\textheight,keepaspectratio]{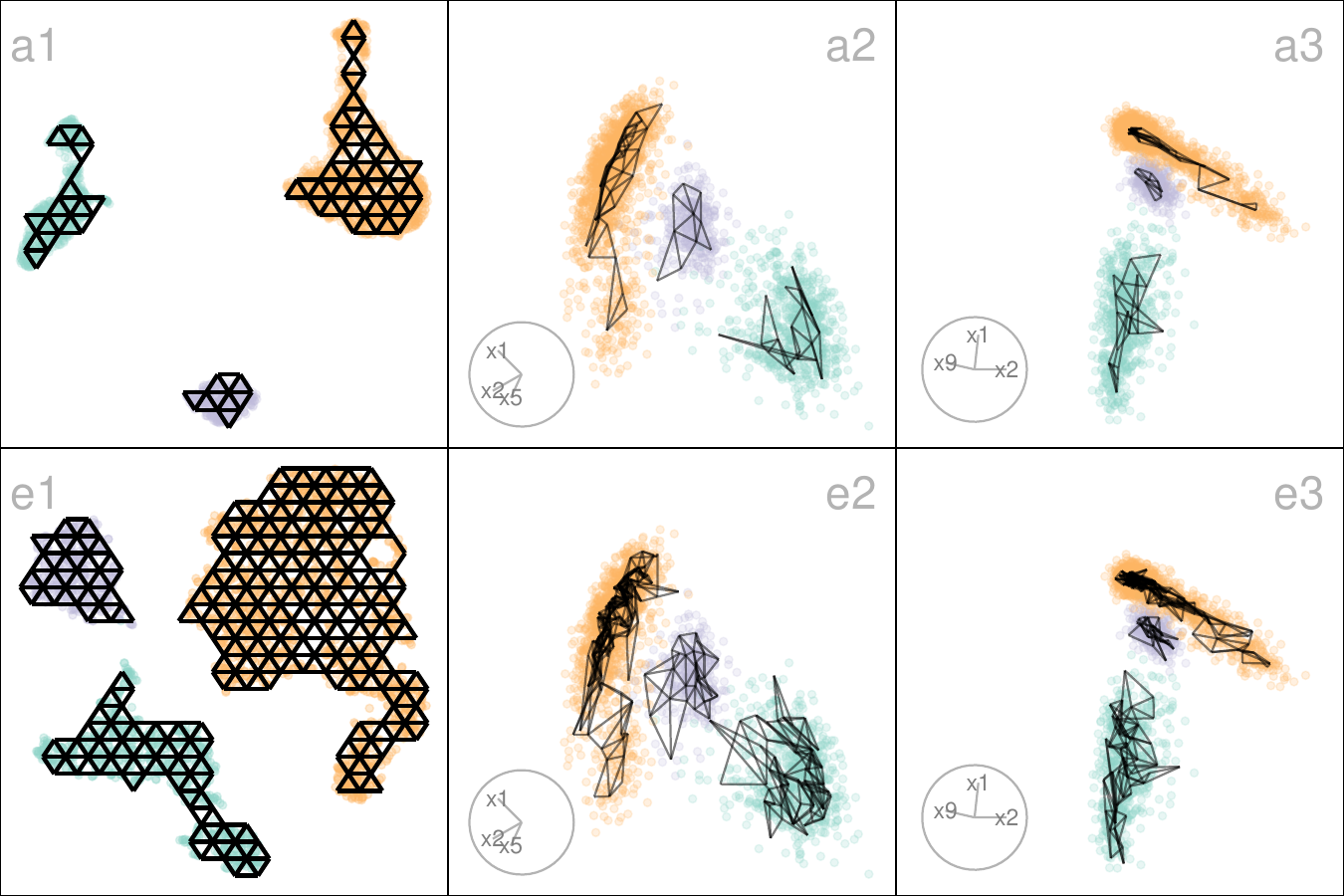}

}

\caption{\label{fig-model-pbmc-author-proj}Compare the published
\(2\text{-}D\) layout (a) made with UMAP and the \(2\text{-}D\) layout
selected by HBE plot (e) made by tSNE. The two plots on the right show
projections from a tour, with the models overlaid. The published layout
a suggested three very separated clusters, but this is not present in
the data. While there may be three clusters they are not well-separated.
The difference in model fit also indicates this: the published layout a
does not spread out fully into the point cloud like the model generated
from layout e. This supports the choice that layout e is the better
representation of the data, because it does not exaggerate separation
between clusters.}

\end{figure}

To further assess the choices, we need to look at the model in the data
space, by using a tour to show the wireframe model overlaid on the data
in the \(9\text{-}D\) space (Figure~\ref{fig-model-pbmc-author-proj}).
Here we compare the published layout (a) versus what we argue is the
best layout (e). The top row (a1, a2, a3) correspond to the published
layout and the bottom row (e1, e2, e3) correspond to the optimal choice
according to our procedure. The middle and right plots show two
projections. The primary difference between the two models is that the
model of layout e does not fill out to the extent of the data but
concentrates in the center of each point cloud. Both suggest that three
clusters is a reasonable interpretation of the structure, but layout e
more accurately reflects the separation between them, which is small.

\subsection{MNIST hand-written digits}\label{sec-mnist}

The digit ``1'' of the MNIST dataset \citep{lecun1998} consists of
\(7877\) grayscale images of handwritten ``1''s. Each image is
\(28 \times 28\) pixels which corresponds to \(784\) variables. The
first \(10\) principal components, explaining \(83\%\) of the total
variation, are used. This data essentially lies on a nonlinear manifold
in the high dimensions, defined by the shapes that ``1''s make when
sketched. We expect that the best layout captures this type of structure
and does not exhibit distinct clusters.

\begin{figure}[!ht]

\centering{

\includegraphics[width=1\linewidth,height=\textheight,keepaspectratio]{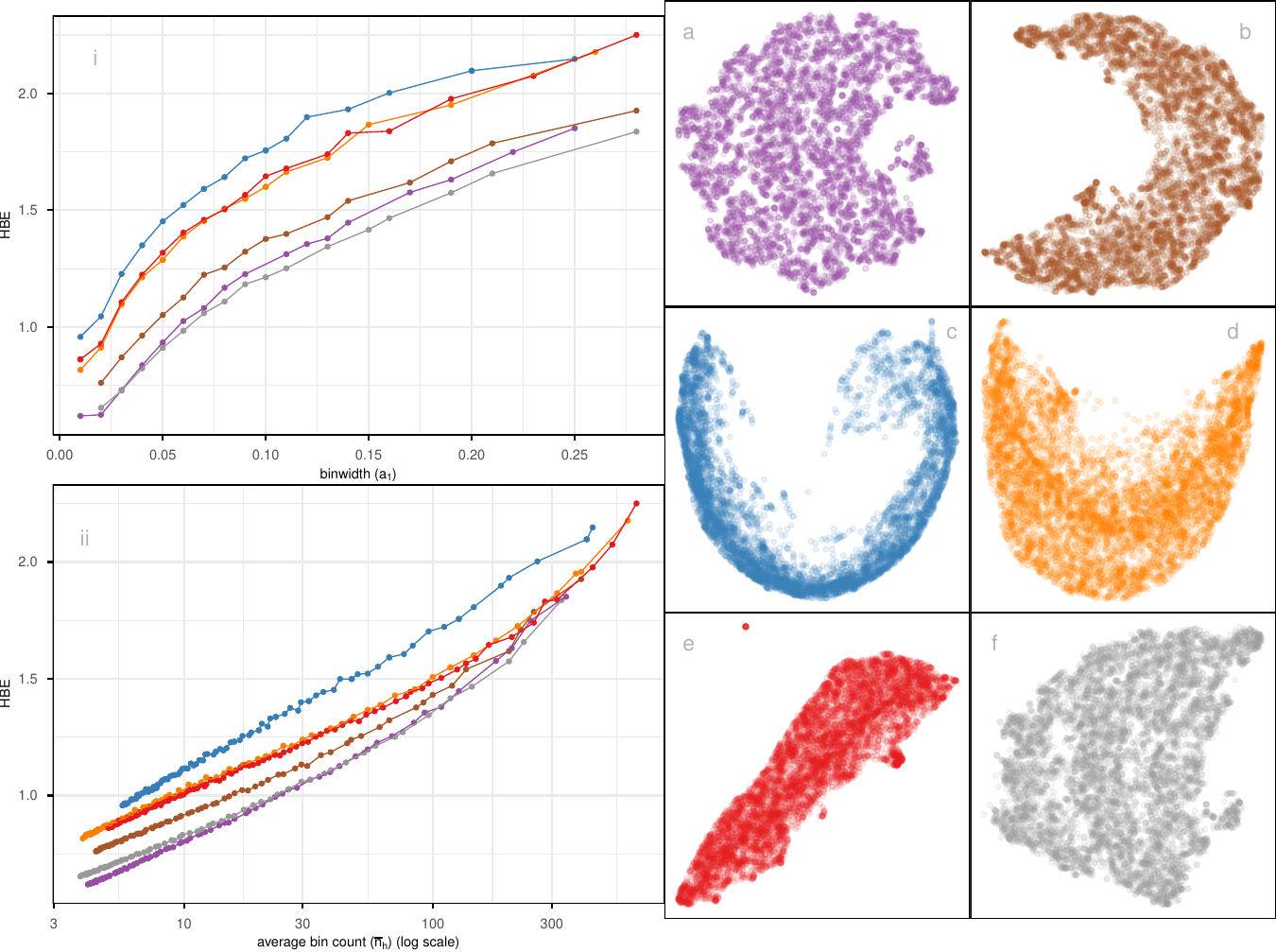}

}

\caption{\label{fig-mnist-rmse}Assessing which of the 6 NLDR layouts of
the MNIST digit 1 data is the better representation using HBE for
varying (i) binwidth (\(a_1\)), and (ii) average bin count
(\(\bar{n}_h\)). Colour is used for the lines and points in the left
plot to match the scatterplots of the NLDR layouts (a-f). Layout c is
universally poor. Layouts a, f that show a big cluster and a small
circular cluster are universally optimal. Layout a performs well at tiny
binwidth (where most points are in their own bin) and not as well as f
with larger binwidth, thus layout f is the best choice. Plot (ii), which
accounts for the density within clusters by using average bin count,
shows reduced differences between layouts, indicating that part of the
variation in (i) is driven by cluster density rather than true
structural differences.}

\end{figure}%

Figure~\ref{fig-mnist-rmse} compares the fit of six layouts computed
using UMAP (b), PHATE (c), TriMAP (d), PaCMAP (e) with default
hyper-parameter setting and two tSNE runs, one with default
hyper-parameter setting (a) and the other changing perplexity to \(89\)
(f). The layouts are reasonably similar in that they all have the
observations in a single blob. Some (b, c) have a more curved shape than
others. Layout e is the most different having a linear shape, and a
single very large outlier. Both a and f have a small clump of points
perhaps slightly disconnected from the other points, in the lower to
middle right.

\begin{figure}[!ht]

\centering{

\includegraphics[width=1\linewidth,height=\textheight,keepaspectratio]{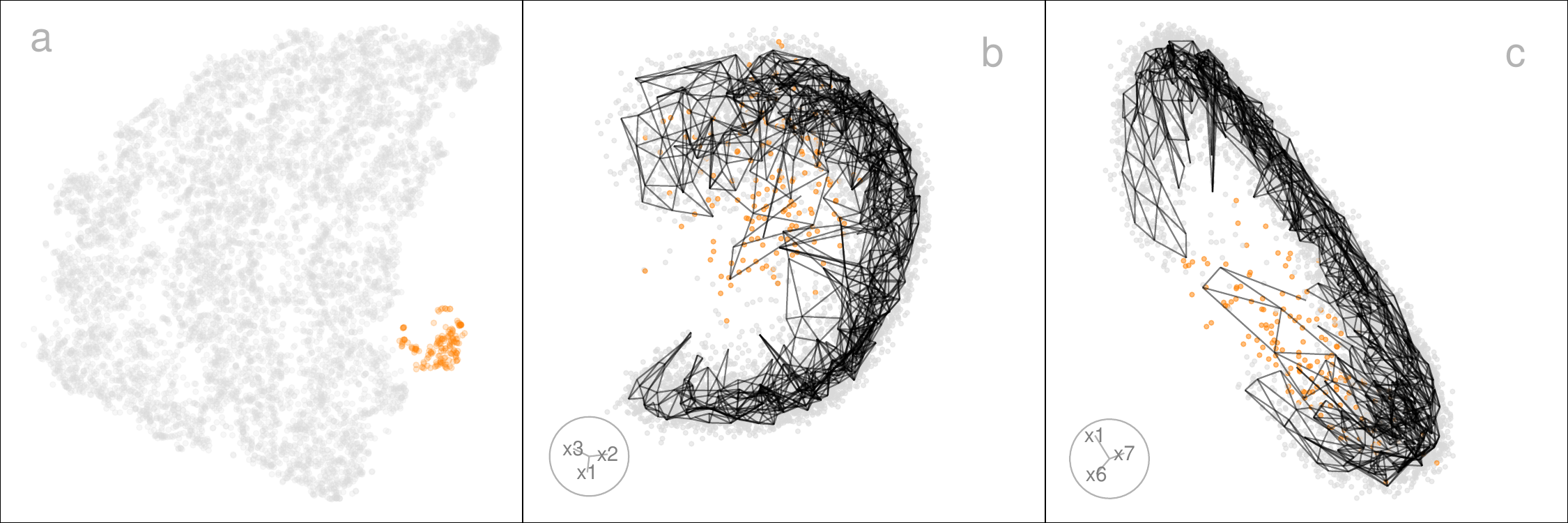}

}

\caption{\label{fig-clust-mnist}Summary from exploring tSNE layout of
the MNIST digit 1 data (a in Figure 12) using linked brushing. There is
a big nonlinear cluster (grey) and a small cluster (orange) located very
close to the one corner of the big cluster in \(2\text{-}D\) (a). The
MNIST digit 1 data has a nonlinear structure in \(10\text{-}D\). Two
\(2\text{-}D\) projections from a tour on \(10\text{-}D\) reveal that
the small orange cluster is actually a diffuse set of points wrapped
within the grey cluster, which is C-shaped in the high dimensions.}

\end{figure}

The layout plots are colored to match the lines in the HBE vs binwidth
(\(a_1\)) plot. Layouts a, b and f fit the data better than c, d, e, and
layout f appears to be the best fit. Figure~\ref{fig-clust-mnist} shows
this model in the data space in two projections from a tour. The data is
curved in the \(10\text{-}D\) space, and the fitted model captures this
curve. The small clump of points in the \(2\text{-}D\) layout is
highlighted in both displays. These are almost all inside the curve of
the bulk of points and are sparsely located. The fact that they are
packed together in the \(2\text{-}D\) layout is likely due to the
handling of density differences by the NLDR.

An interesting aside is that the rather strange layout e, which has what
looks like a single point far from the remaining observations is
actually similar to this one. That point is actually a clump of points
corresponding to some of the diffuse points interior to the curve of the
bulk of points. This is easy to see using the linked brushing tool.

\begin{figure}[!ht]

\centering{

\includegraphics[width=1\linewidth,height=\textheight,keepaspectratio]{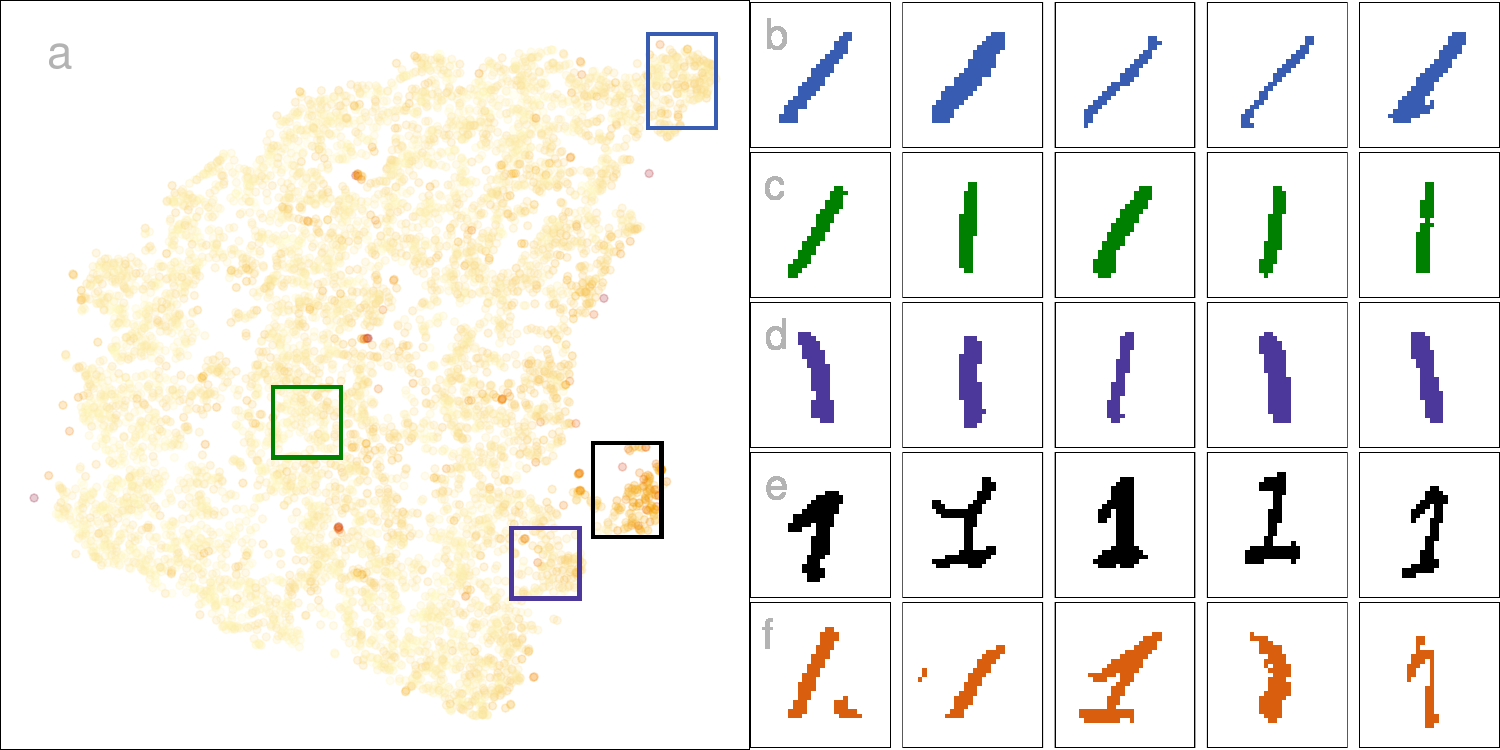}

}

\caption{\label{fig-model-error-mnist}Summary of the layout structure,
and large errors, relative to the MNIST digit 1 shape: (a) layout
colored by residual value, and at right (b-e) are images of samples of
observations taken at locations around the layout, showing similarity in
how the 1's were drawn. Set (f) are images corresponding to large
residuals in the big cluster (darker orange in plot a). Along the big
cluster, the angle of digit 1 changes (b-d). The small cluster has
larger residuals, and the images show that these tend to be European
style with a flag at the top, and a base at the bottom. The set in (f)
show various poorly written digits.}

\end{figure}

The next step is to investigate the \(2\text{-}D\) layout to understand
what information is learned from this representation.
Figure~\ref{fig-model-error-mnist} summarizes this investigation. Plot a
shows the layout with points colored by their residual value - darker
color indicates larger residual and poor fit. The plots b, c, d, e show
samples of hand-written digits taken from inside the colored boxes.
Going from top to bottom around the curve shape we can see that the
``1''s are drawn with from right slant to a left slant. The ``1''s in d
(black box) tend to have the extra up stroke but are quite varied in
appearance. The ``1''s shown in the plots labelled e correspond to
points with big residuals. They can be seen to be more strangely drawn
than the others. Overall, this \(2\text{-}D\) layout shows a useful way
to summarize the variation in way ``1''s are drawn.

\section{Discussion}\label{sec-discussion-visalgo}

We have developed an approach to help assess and compare NLDR layouts,
generated by different methods and hyper-parameter choice(s). It depends
on conceptualizing the \(2\text{-}D\) layout as a model, allowing for
the creation of a wireframe representation of the model that can be
lifted into \(p\text{-}D\). The fit is assessed by viewing the model in
the data space, computing residuals and HBE. Different layouts can be
compared using the HBE, providing a quantitative metric to decide on the
most suitable NLDR layout to represent the \(p\text{-}D\) data. Global
and local preservation of structure is assessed by examining the HBE
across a range of binwidths. It also provides a way to predict the
values of new \(p\text{-}D\) observations in the \(2\text{-}D\), which
could be useful for implementing uncertainty checks, such as using
training and testing samples.

The new methodology is accompanied by an R package called
\texttt{quollr}, so that it is readily usable and broadly accessible.
The package has methods to fit the model, compute diagnostics and also
visualize the results, with interactivity. We have primarily used the
\texttt{langevitour} software \citep{harisson2024} to view the model in
the data space, but other tour software such as \texttt{tourr}
\citep{wickham2011} and \texttt{detourr} \citep{hart2022} could also be
used.

Two examples illustrating usage are provided: the PBMC3k data, where the
NLDR is summarizing clustering in \(p\text{-}D\) and hand-written digits
illustrating how NLDR represents an intrinsically low-dimensional
nonlinear manifold. We examined a typical published usage of UMAP with
the PBMC3k dataset \citep{chen2024}. As is typical of UMAP layout with
default settings, the separation between clusters is grossly
exaggerated. The layout even suggests separation where there is none.
Our approach provides a way to choose a reasonable layout and avoids the
use of misleading layouts in the future. In the hand-written digits
\citep{lecun1998}, we illustrate how our model fit statistics show that
a flat disc layout is superior to the curved-shaped layouts, and how to
identify oddly written ``1''s using the residuals of the fitted model.

This work can be applied with existing metrics for evaluating NLDR
layout, such as ARNX, RTA, SC, and RGS. It provides an additional
evaluation metric, and importantly allows any layout to be viewed in the
\(p\text{-}D\) data space. This latter aspect can help to disentangle
conflicting suggestions by the different metrics.

Additional exploration of distance measures to summarize the fit could
be a valuable direction for future work. We have used Euclidean
distance, but other measures, such as geodesic distances
\citep{joshua2000}, may better capture curved or nonlinear relationships
in the data and are worth exploring.

This work has also revealed some interesting behaviors of NLDR methods,
including twisting, flattened ``pancake'' clusters in \(p\text{-}D\),
and severe effects of density differences. These are described in more
detail in the supplementary materials.

Researchers usually use \(2\text{-}D\) layouts, but if a \(k\text{-}D\)
(\(k>2\)) layout is provided the approach developed here could be
extended. Potential approaches include \(3\text{-}D\) binning,
\(k\)-means clustering, or even special implementations of convex hulls.

\section{Supplementary Materials}\label{sec-supplementary}

All the materials to reproduce the paper can be found at
\url{https://github.com/JayaniLakshika/paper-nldr-vis-algorithm}.

The supplementary materials provide additional details on the methods
and hyper-parameters used to generate layouts, video links of animated
\(p\text{-}D\) tours, notation summaries, and the R and Python scripts
used in the study. They also describe the generation of the 2NC7 data,
computation of hexagon grid configurations, and data binning procedures.
Further sections highlight interesting NLDR behaviors observed in the
data space and compare HBE with existing evaluation metrics for the
PBMC3k and MNIST datasets.

The R package \texttt{quollr}, available on CRAN and at
\url{https://jayanilakshika.github.io/quollr/}, provides software
accompaying this paper to fit the wireframe model representation,
compute diagnostics, visualize the model in the data with
\texttt{langevitour} and link multiple plots interactively.

\section{Acknowledgments}\label{acknowledgments}

These \texttt{R} packages were used for the work: \texttt{tidyverse}
\citep{hadley2019}, \texttt{Rtsne} \citep{jesse2015}, \texttt{umap}
\citep{tomasz2023}, \texttt{patchwork} \citep{thomas2024},
\texttt{colorspace} \citep{achim2020}, \texttt{langevitour}
\citep{harisson2024}, \texttt{conflicted} \citep{hadley2023},
\texttt{reticulate} \citep{kevin2024}, \texttt{kableExtra}
\citep{hao2024}. These \texttt{python} packages were used for the work:
\texttt{trimap} \citep{amid2022} and \texttt{pacmap}
\citep{yingfan2021}. The article was created with \texttt{R} packages
\texttt{quarto} \citep{jjallaire2024}.

\section*{References}\label{references}
\addcontentsline{toc}{section}{References}

\renewcommand{\bibsection}{}
\bibliography{paper.bib}

\end{document}